\documentstyle[twoside]{article}
\input{psfig}

\catcode`\@=11
\long\def\@makefntext#1{
\protect\noindent \hbox to 3.2pt {\hskip-.9pt  
$^{{\eightrm\@thefnmark}}$\hfil}#1\hfill}		%CAN BE USED 

\def\thefootnote{\fnsymbol{footnote}}
\def\@makefnmark{\hbox to 0pt{$^{\@thefnmark}$\hss}}	%ORIGINAL 
	
\def\ps@myheadings{\let\@mkboth\@gobbletwo
\def\@oddhead{\hbox{}
\rightmark\hfil\eightrm\thepage}   
\def\@oddfoot{}\def\@evenhead{\eightrm\thepage\hfil
\leftmark\hbox{}}\def\@evenfoot{}
\def\sectionmark##1{}\def\subsectionmark##1{}}

\oddsidemargin=\evensidemargin
\renewcommand{\thefootnote}{\fnsymbol{footnote}}

\newcounter{sectionc}\newcounter{subsectionc}\newcounter{subsubsectionc}
\renewcommand{\section}[1] {\vspace{12pt}\addtocounter{sectionc}{1} 
\setcounter{subsectionc}{0}\setcounter{subsubsectionc}{0}\noindent 
	{\tenbf\thesectionc. #1}\par\vspace{5pt}}
\renewcommand{\subsection}[1] {\vspace{12pt}\addtocounter{subsectionc}{1} 
	\setcounter{subsubsectionc}{0}\noindent 
	{\bf\thesectionc.\thesubsectionc. {\kern1pt \bfit #1}}\par\vspace{5pt}}
\renewcommand{\subsubsection}[1] {\vspace{12pt}\addtocounter{subsubsectionc}{1}
	\noindent{\tenrm\thesectionc.\thesubsectionc.\thesubsubsectionc.
	{\kern1pt \tenit #1}}\par\vspace{5pt}}
\newcommand{\nonumsection}[1] {\vspace{12pt}\noindent{\tenbf #1}
	\par\vspace{5pt}}

\newcounter{appendixc}
\newcounter{subappendixc}[appendixc]
\newcounter{subsubappendixc}[subappendixc]
\renewcommand{\thesubappendixc}{\Alph{appendixc}.\arabic{subappendixc}}
\renewcommand{\thesubsubappendixc}
	{\Alph{appendixc}.\arabic{subappendixc}.\arabic{subsubappendixc}}

\renewcommand{\appendix}[1] {\vspace{12pt}
        \refstepcounter{appendixc}
        \setcounter{figure}{0}
        \setcounter{table}{0}
        \setcounter{lemma}{0}
        \setcounter{theorem}{0}
        \setcounter{corollary}{0}
        \setcounter{definition}{0}
        \setcounter{equation}{0}
        \renewcommand{\thefigure}{\Alph{appendixc}.\arabic{figure}}
        \renewcommand{\thetable}{\Alph{appendixc}.\arabic{table}}
        \renewcommand{\theappendixc}{\Alph{appendixc}}
        \renewcommand{\thelemma}{\Alph{appendixc}.\arabic{lemma}}
        \renewcommand{\thetheorem}{\Alph{appendixc}.\arabic{theorem}}
        \renewcommand{\thedefinition}{\Alph{appendixc}.\arabic{definition}}
        \renewcommand{\thecorollary}{\Alph{appendixc}.\arabic{corollary}}
        \renewcommand{\theequation}{\Alph{appendixc}.\arabic{equation}}
        \noindent{\tenbf Appendix \theappendixc #1}\par\vspace{5pt}}
\newcommand{\subappendix}[1] {\vspace{12pt}
        \refstepcounter{subappendixc}
        \noindent{\bf Appendix \thesubappendixc. {\kern1pt \bfit #1}}
	\par\vspace{5pt}}
\newcommand{\subsubappendix}[1] {\vspace{12pt}
        \refstepcounter{subsubappendixc}
        \noindent{\rm Appendix \thesubsubappendixc. {\kern1pt \tenit #1}}
	\par\vspace{5pt}}

\newcommand{\textlineskip}{\baselineskip=13pt}
\newcommand{\smalllineskip}{\baselineskip=10pt}

\def\eightcirc{
\begin{picture}(0,0)
\put(4.4,1.8){\circle{6.5}}
\end{picture}}
\def\eightcopyright{\eightcirc\kern2.7pt\hbox{\eightrm c}} 

\newcommand{\copyrightheading}[1]
	{\vspace*{-2.5cm}\smalllineskip{\flushleft
	{\footnotesize International Journal of Modern Physics A, #1}\\
	{\footnotesize $\eightcopyright$\, World Scientific Publishing
	 Company}\\
	 }}

\def\abstracts#1#2#3{{
	\centering{\begin{minipage}{4.5in}\baselineskip=10pt\footnotesize
	\parindent=0pt #1\par 
	\parindent=15pt #2\par
	\parindent=15pt #3
	\end{minipage}}\par}} 

\newcommand{\bibit}{\nineit}

\renewenvironment{thebibliography}[1]
	{\frenchspacing
	 \ninerm\baselineskip=11pt
	 \begin{list}{\arabic{enumi}.}
	{\usecounter{enumi}\setlength{\parsep}{0pt}
	 \setlength{\leftmargin 12.7pt}{\rightmargin 0pt} %FOR 1--9 ITEMS
	 \setlength{\itemsep}{0pt} \settowidth
	{\labelwidth}{#1.}\sloppy}}{\end{list}}

\newcounter{itemlistc}
\newcounter{romanlistc}
\newcounter{alphlistc}
\newcounter{arabiclistc}

\newcommand{\fcaption}[1]{
        \refstepcounter{figure}
        \setbox\@tempboxa = \hbox{\footnotesize Fig.~\thefigure. #1}
        \ifdim \wd\@tempboxa > 5in
           {\begin{center}
        \parbox{5in}{\footnotesize\smalllineskip Fig.~\thefigure. #1}
            \end{center}}
        \else
             {\begin{center}
             {\footnotesize Fig.~\thefigure. #1}
              \end{center}}
        \fi}

\newcommand{\tcaption}[1]{
        \refstepcounter{table}
        \setbox\@tempboxa = \hbox{\footnotesize Table~\thetable. #1}
        \ifdim \wd\@tempboxa > 5in
           {\begin{center}
        \parbox{5in}{\footnotesize\smalllineskip Table~\thetable. #1}
            \end{center}}
        \else
             {\begin{center}
             {\footnotesize Table~\thetable. #1}
              \end{center}}
        \fi}

\def\@citex[#1]#2{\if@filesw\immediate\write\@auxout
	{\string\citation{#2}}\fi
\def\@citea{}\@cite{\@for\@citeb:=#2\do
	{\@citea\def\@citea{,}\@ifundefined
	{b@\@citeb}{{\bf ?}\@warning
	{Citation `\@citeb' on page \thepage \space undefined}}
	{\csname b@\@citeb\endcsname}}}{#1}}

\newif\if@cghi
\def\cite{\@cghitrue\@ifnextchar [{\@tempswatrue
	\@citex}{\@tempswafalse\@citex[]}}
\def\citelow{\@cghifalse\@ifnextchar [{\@tempswatrue
	\@citex}{\@tempswafalse\@citex[]}}
\def\@cite#1#2{{$\null^{#1}$\if@tempswa\typeout
	{IJCGA warning: optional citation argument 
	ignored: `#2'} \fi}}

\def\pmb#1{\setbox0=\hbox{#1}
	\kern-.025em\copy0\kern-\wd0
	\kern.05em\copy0\kern-\wd0
	\kern-.025em\raise.0433em\box0}

\def\fnt#1#2{\footnotetext{\kern-.3em
	{$^{\mbox{\scriptsize #1}}$}{#2}}}

\def\fpage#1{\begingroup
\voffset=.3in
\thispagestyle{empty}\begin{table}[b]\centerline{\footnotesize #1}
	\end{table}\endgroup}

\font\tenrm=cmr10
\font\tenit=cmti10 
\font\tenbf=cmbx10
\font\bfit=cmbxti10 at 10pt
\font\ninerm=cmr9
\font\nineit=cmti9

\font\eightrm=cmr8

\def\qed{\hbox{${\vcenter{\vbox{			%HOLLOW SQUARE
   \hrule height 0.4pt\hbox{\vrule width 0.4pt height 6pt
   \kern5pt\vrule width 0.4pt}\hrule height 0.4pt}}}$}}

\renewcommand{\thefootnote}{\fnsymbol{footnote}}	%USE SYMBOLIC FOOTNOTE

\begin{document}

%\runninghead{Instructions for Typesetting Camera-Ready
%Manuscripts $\ldots$} {Instructions for Typesetting Camera-Ready
%Manuscripts $\ldots$}

\normalsize\textlineskip
\thispagestyle{empty}
\setcounter{page}{1}

\copyrightheading{}			%{Vol. 0, No. 0 (1993) 000--000}

\vspace*{0.88truein}

\fpage{1}
\centerline{\bf NEW DIRECTIONS IN DATA ANALYSIS\footnote{Talk given at the DPF2000 meeting in Columbus, Ohio.}}
\vspace*{0.035truein}
%\centerline{\bf  
%}
\vspace*{0.37truein}
\centerline{\footnotesize PUSHPALATHA C. BHAT}
\vspace*{0.015truein}
\centerline{\footnotesize\it 
Fermi National Accelerator Laboratory\footnote{
Operated by the Universities Research Association, Inc., under
contract with the U.\ S.\ Department of Energy.} 
, P.O. Box 500
}
\baselineskip=10pt
\centerline{\footnotesize\it Batavia, Illinois 60510, USA}

%\vspace*{0.225truein}
%\publisher{(received date)}{(revised date)}

\vspace*{0.21truein}    \abstracts{  In the next decade, high
energy physicists will  use  very sophisticated equipment   to record
unprecedented amounts of data in the hope  of making major advances in
our understanding of  particle phenomena.  Some of  the signals of new
physics  will be small, and  the  use of advanced analysis  techniques
will be crucial for optimizing signal to noise ratio. I will discuss 
new directions in data analysis and some
novel methods that could prove to be particularly valuable
for finding   evidence of any   new physics, for   improving precision
measurements and for exploring parameter  spaces of theoretical models.
  }{}{} 

%\textlineskip			%) USE THIS MEASUREMENT WHEN THERE IS
%\vspace*{12pt}			%) NO SECTION HEADING

\vspace*{1pt}\textlineskip %) USE THIS MEASUREMENT WHEN THERE IS 
\section{Introduction}	%) A SECTION HEADING
\vspace*{-0.5pt} 
\noindent
We are building powerful accelerators  and sophisticated detectors  in
the hope of making major discoveries in the next decade.  We hope
to discover the Higgs boson, Supersymmetry or Technicolor or something
completely unexpected and exalting!     We  would also like  to   make
precision measurements of some of  nature's fundamental parameters. In
order to achieve   these  goals  it  is crucial that  we employ
advanced and optimal data analysis methods both on-line and off-line.

In the not so distant past, we could afford the  luxury of writing 
data onto storage media with simple interaction triggers and organize,
reduce and analyze data completely off-line.   But,  as we learnt more
about  the world and  began to address  more complex problems, looking
for extremely rare   processes  at higher  beam  energies and   higher
luminosities, we had to handle and  sift through large amounts of data
on-line before selected  data are written  out. The new  generation of
experiments will be a lot  more demanding than   the previous in  data
handling at all  stages; the rates of  interactions to be  handled and
the number  of  detector channels read-out  will  grow by  orders  of
magnitude.  Finding the signals of new physics becomes a veritable case
of ``finding  needles  in a   hay-stack''.  The unprecedented challenges
will require  new paradigms and  technologies to be identified, 
developed and adopted.

%\pagebreak 

\textheight=7.8truein                         \setcounter{footnote}{0}
\renewcommand{\thefootnote}{\alph{footnote}} 

\section{Intelligent Detectors and Smart Triggers}
\noindent
The data analysis  in HEP experiments starts  when a high energy event
occurs. The data from the detectors must  be transformed into useful
``physics'' information in real-time.   The calorimeter, for instance,
can have ``intelligence'' close to its electronic read-out so that the
clustering  and  energy   measurements  are   readily  available. Such
information from different sub-detectors can be  used to extract event
features,  such as the number  of tracks, high transverse momentum ($p_T$) 
objects and object identities.  These features can then be used to
make a  global decision about  whether or not the event is potentially
interesting. Therefore, we need to build 
{\it intelligent detectors}   and 
{\it triggers}. The feature  extraction and
further processing  such as  particle identification or event classification
can be accomplished using  smart algorithms either built into hardware
(neural networks  chips,  for example) or  configured  in generic
hardware such  as Field  Programmable  Gate Arrays  (FPGAs) or Digital
Signal Processors (DSPs). The  H1 experiment  at HERA has  implemented
neural network hardware  in  its Level-2 trigger$^1$ which has 
operated  successfully since 1996  and  has been  crucial for the rich
harvest of physics results from H1.  Innovative   data management on-line,  
using for example RAM   disks, and employing  algorithms  in  trigger
hardware would be beneficial  in meeting the  demands of data handling
and  analysis on-line.   Use   of expert or fuzzy-logic  systems in
controls  and monitoring of   detector electronics is  an area that
needs to be explored as well.

\section{Optimal Analysis Methods}
\noindent
When   classifying    events, the    traditional  procedure   of
choosing and applying cuts on one   event variable at a time is
rarely  optimal  in  the  sense  of  minimizing   the  probability  to
mis-classify events.  By contrast, given a set of event variables,
that, in general, are correlated, optimal separation  can
always be achieved if one treats the variables in a fully {\it multivariate}
manner.  Some  classes  of   data   analysis  tasks that    benefit from
multivariate methods are  particle identification (electrons, taus,
$b$-quark jets    etc.,   $\gamma  /     \pi^0$  separation,   quark/gluon  
jet separation), signal-background   discrimination,  parameter  estimation
(tracking, vertexing, mass measurements etc), function approximation (
energy correction  functions, mis-identification rates) and data exploration
(latent structure analysis, multivariate bump hunting).

There  are a  number   of parametric and non-parametric   multivariate
methods from  the simplest Fisher  linear discriminant to the
sophisticated  non-linear,    adaptive  methods$^2$.  Neural
networks   have  emerged  as    powerful    and flexible methods   of
multivariate data analysis. These and  other multivariate methods have been
used in recent experimental data analyses  around the world: D\O,
CDF at   the  Tevatron,  LEP  experiments at CERN,  experiments  at   DESY and
SLC. These will be  the methods of choice for future  analyses.

\section{ Some Examples }
\noindent
Because of a lack of  space, I
describe a couple of  example analyses  from the  D\O\ experiment  
and a Monte Carlo study  (with apologies   to other  experiments).

The top  quark mass measurement  was  one  of  the most important  
results from the last run of the  Tevatron experiments$^3$.  Since  the D\O\
experiment did not have a silicon vertex  detector (SVX) in Run I and
used only soft muon tagging for b-jet identification, 
the  b-tagging efficiency was only 20\% in the
lepton $+\ge$~4jets channel compared to approximately  53\% at CDF which
had the  ability to tag b-jets with  its SVX.  Nonetheless, D\O\ was
able to  measure the top quark  mass with a precision  approaching that of CDF,
by using multivariate techniques  for separating signal and background
while  minimizing the correlation of  the selection with the top quark
mass.   Two  multivariate methods,  (1)~a log-likelihood technique and
(2)~a feed   forward neural network, were used   to  compute a signal
probability $P(top\mid D)$ for each event, given data $D$.  A likelihood
fit (based on a Bayesian method$^4$) of the data to a discrete set of
signal and background models    in the [$P(top \mid D), m_{fit}$]   plane  was
used to extract  the top quark mass.  Combining  results from the
two methods, taking into account  their  correlation, an overall
measurement  of  $m_t=173.3\pm7.8$~GeV/c$^2$ was  obtained.   The top quark
mass  measurement in the  dilepton+jet  channel also  benefitted from
a multivariate method (probability density estimator). 

Multivariate analysis methods 
enabled D\O\ to establish the  world's best limits$^5$ on the existence
of leptoquarks that might decay into electrons  and quarks.  The lower
limit   on the leptoquark mass  from  the D\O\  experiment  was 225 GeV,
ruling  out the possibility that   the HERA  event excess (reported  in
February 1997) can be interpreted  as an evidence for first  generation
scalar leptoquark production. 

In the next few years, a low mass Higgs boson may be discovered 
at the Fermilab Tevatron, a possibility that has motivated an intense 
study$^6$.  We have studied$^6$ the potential of
 the  CDF and D\O\  experiments to make such a  discovery in Run II, via
 the  processes  $p\bar{p}\rightarrow WH  \rightarrow  l\nu b\bar{b}$,
 $p\bar{p}\rightarrow    ZH\rightarrow    l^+l^-      b\bar{b}$    and
 $p\bar{p}\rightarrow ZH\rightarrow \nu \bar{\nu} b\bar{b}$.  We have
 shown that a neural network analysis could yield a 5$\sigma$  discovery
 for $100\le M_H\le  130$GeV/c$^2$ with only half the integrated
 luminosity needed for a conventional analysis.  Fig. 1  shows
 the neural  network distributions for  signal monte  carlo events with
 $M_H=110$ GeV/c$^2$ compared with the  specified backgrounds, for a set
of seven input variables. (  For
 details, see  ref. 6). A plot  of  the required integrated luminosity
 for a 5$\sigma$ observation is also shown in Fig 1. 

\section{Exploring Models}
\noindent
Physicists  are becoming  increasingly  convinced of the
value of Bayesian reasoning as a  powerful way of extracting information
from data  and of updating our knowledge  upon  arrival of   new  data.
The Bayesian approach provides  a well-founded mathematical procedure to
compute   the  conditional probability  of a   model  and therefore to
do straight-forward and meaningful        model
comparisons.  It  also allows  treatment  of  all uncertainties  in  a
consistent manner.  Practical applications of these ideas to (1)fitting
binned data to one or more multi-source models$^4$ and (2)the extraction
of the solar  neutrino survival probability$^7$  using data and solar neutrino
model predictions illustrate  the usefulness of  Bayesian methods
in data analysis. 

This     approach provides a systematic way of extracting
probabilistic   information for each  parameter  of a  model,  say for
example a particular   SUSY model,   via  marginalization over  the
remaining parameters.  I believe that this probabilistic approach to model
exploration could prove to be extremely fruitful.  Studies of this approach are 
in progress.

\begin{figure}[htbp]
\vspace*{5pt}
\centering
    \begin{minipage}[t]{.45\linewidth}
    \hspace*{-8pt}
    \mbox{\psfig{figure=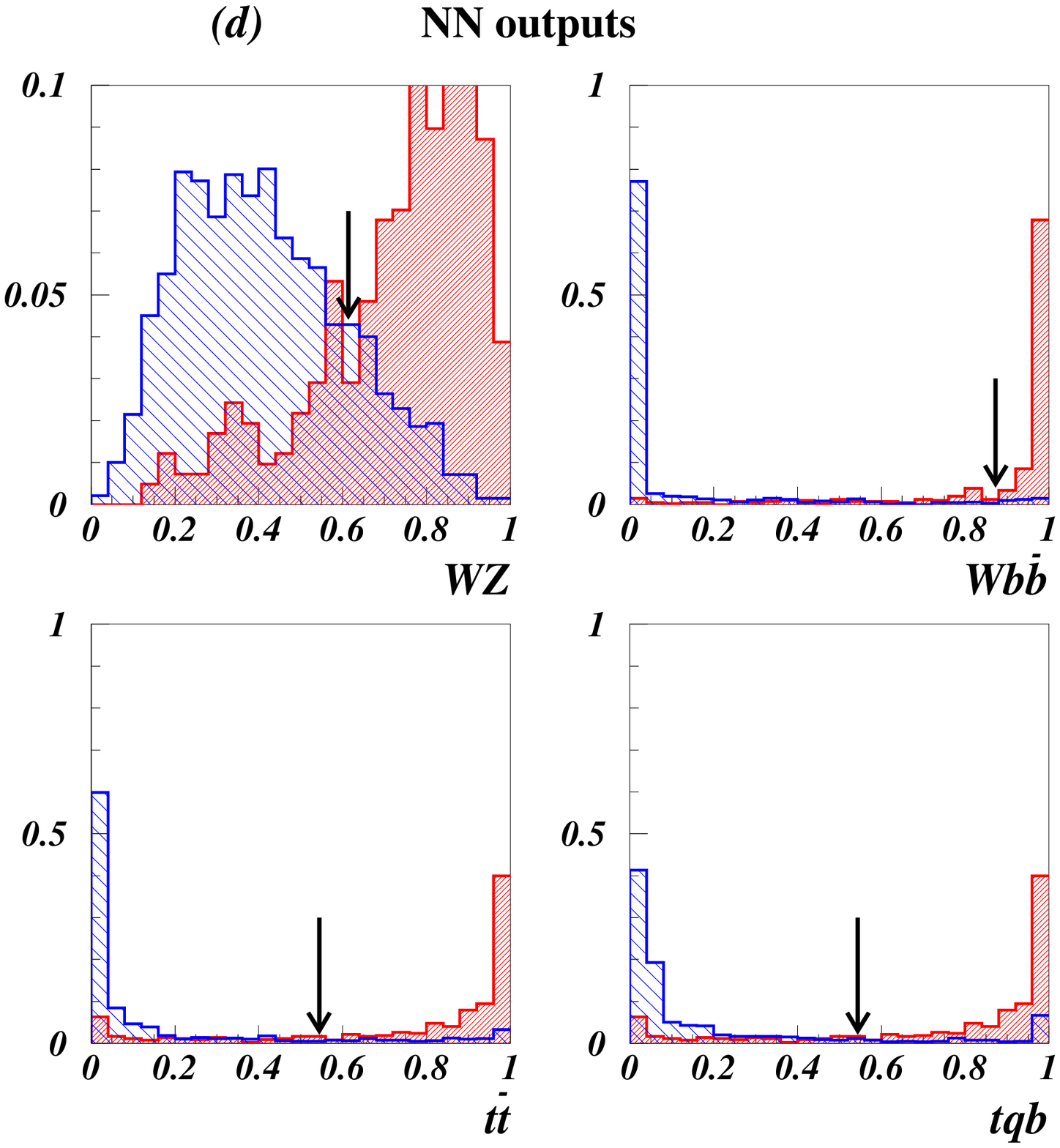,width=7cm,height=5cm}}
    \end{minipage}
    \begin{minipage}[t]{.45\linewidth}
    \hspace*{17pt}
    \mbox{\psfig{figure=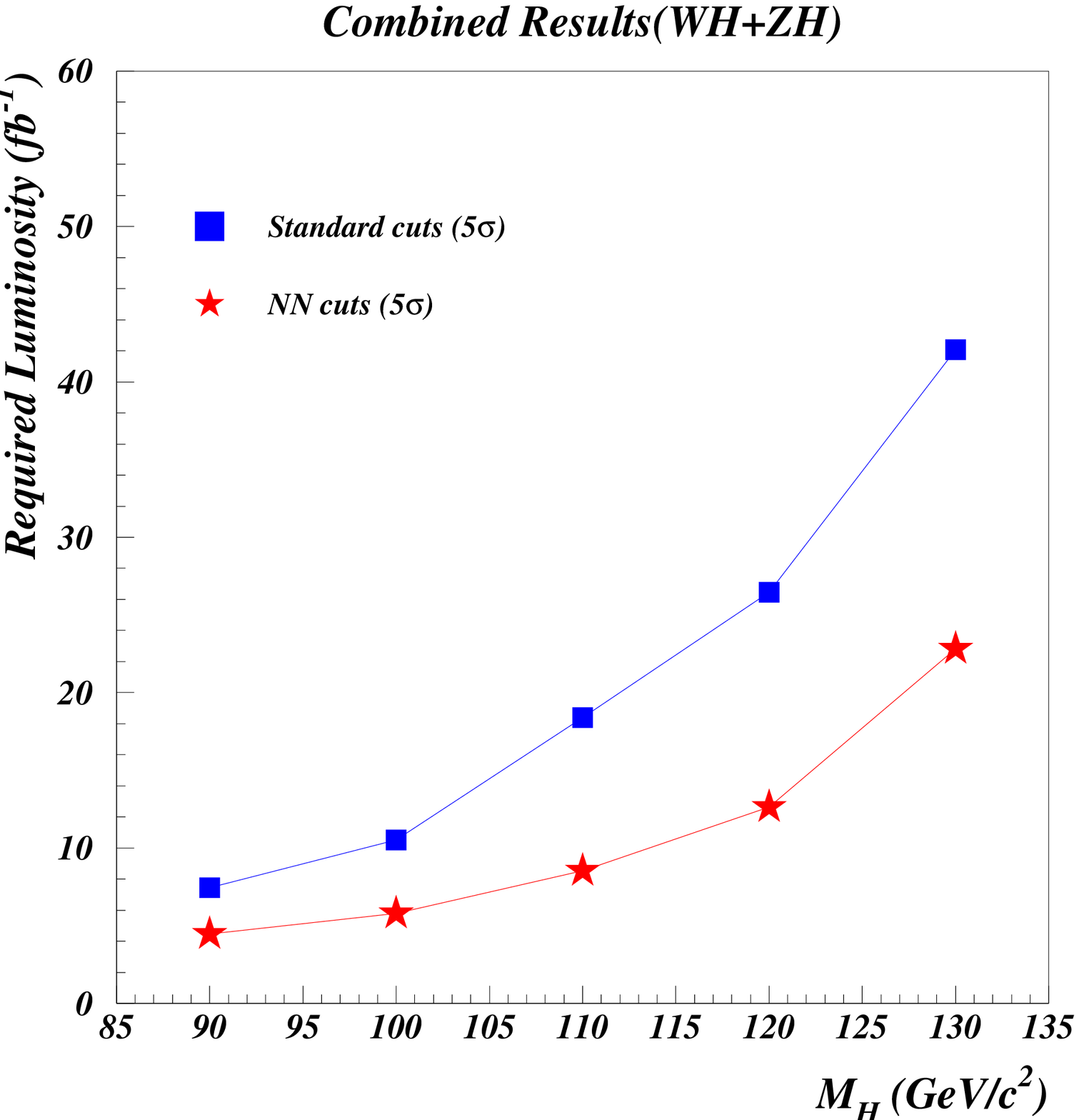,width=5.5cm,height=5cm}}
    \end{minipage}
%\psfig{figure=net_p000.eps,width=5cm,height=5cm}
%\psfig{figure=wh_8.eps,width=5cm,height=5cm}
%\centerline{\vbox{file=net_p000.eps width 5cm height 5cm }}
%\vspace*{1.4truein}		%ORIGINAL SIZE=1.6TRUEIN x 100% - 0.2TRUEIN
%\centerline{\vbox{\hrule width 5cm height0.001pt}}
%\vspace*{13pt}
\fcaption{(Left)The neural network distributions for WH ($M_H=110$ GeV/c$^2$)
 signal Monte Carlo events (heavily shaded) compared with $Wb\bar{b}$, $WZ$,
$t\bar{t}$, single top background events.  (Right) Comparison of the
required integrated luminosities for a 5$\sigma$ observation in the
CDF and D\O\ experiments with $WH$ and $ZH$ channels combined for NN and
conventional cuts.(See ref. 6)}
\end{figure}
\vspace{-0.4cm}
\section{Summary}
\noindent
Major discoveries may await us using the next generation of experiments.
Use of  advanced {\it optimal} analysis methods
will be crucial to discover and study  new physics.  Multivariate methods,
particularly  neural  network techniques have  already had an impact on
discoveries and  precision measurements  and will  be  the methods  of
choice in future analyses.  We have only  scratched the surface in our
use of powerful, multivariate methods  for data  exploration,
visualization and  statistical analysis.  I believe that hybrid methods
combining   ``intelligent''  algorithms   and  the Bayesian/probabilistic
approach will be the wave of the future.

\nonumsection{Acknowledgements} 
\noindent
 The author would  like to thank  Tom Ferbel for proposing this talk and Harrison Prosper for a delightful
collaboration on many of the pursuits discussed  in this paper. 

\nonumsection{References}

\end{document}